\documentclass[12pt]{article}
\usepackage{graphicx}
\usepackage{dcolumn}
\usepackage{bm}
\usepackage{amssymb}
\usepackage{amsmath}
\newcommand{\bq}{\begin{equation}}
\newcommand{\eq}{\end{equation}}

\newcommand{\bqr}{\begin{eqnarray}}
\newcommand{\eqr}{\end{eqnarray}}

\newcommand{\bqrx}{\begin{eqnarray*}}
\newcommand{\eqrx}{\end{eqnarray*}}

\newcommand{\br}{\begin{array}}
\newcommand{\er}{\end{array}}

\newcommand{\qed}{\hspace*{10pt}\rule{1.5ex}{1.5ex}}

\begin{document}

\pagestyle{empty}

\setlength{\parindent}{18pt}
\setlength{\footskip}{.5in}



\vspace*{.6in}
\begin{center}

Unitary conjugation channels with continuous random phases
\end{center}
\begin{center}
Chaobin\, Liu\footnote{cliu@bowiestate.edu} \\Department of Mathematics, Bowie State University \, Bowie, MD 20715 USA\\ 

\end{center}

\begin{abstract}

We consider unitary conjugation channels with continuous random phases. The spectral properties of the channel average are examined, thereby the asymptotic behaviors of the repeated quantum interactions of the motion are derived. We then study the channels with uniformly distributed continuous phases on an interval. In this context, it is shown that discrete phases are sufficient to achieve the channel average.

\end{abstract}

\vskip 0.2in


There have been some recent studies on effects of the randomness imposed on certain quantum scheme models, such as the quantum walks driven by random coin operators determined by i.i.d. phases carried by all matrix elements of the coin operator \cite{KBH2006, JM2010}, and the random quantum channels with randomness defined by Haar-distributed unitaries \cite{HW2008, H2009, CN2010, NP2012}. Motivated by these developments in understanding how the randomness affects the dynamics of the underlying quantum systems, in this work, we study random quantum channels with randomness defined by random diagonal unitary matrices. The spectral properties of this type of channel will be examined with focus on its invariant states. We also discuss how to discretize the random quantum channels in certain specified cases. 

\vskip 0.2 in

In quantum information theory, unitary conjugation channels governing evolution of a closed quantum system, can be written as

$$\Phi_U({\rho})=U\rho U^{\star}$$

where $U$ is the unitary operator (matrix), acting on $\mathbb{C}^d$, and $\rho$ is the density matrix of the quantum system. 

\vskip 0.2 in

We now consider a natural setting of random environment to which a quantum system is exposed, the property of randomness is modeled by so-called i.i.d phases arranged as elements of the governing unitary matrix. In this scenario, the transition of this open quantum system may be modeled by 

$$\rho^{\prime}=\Phi_{U_\theta}(\rho)=U_\theta\rho{U_\theta}^{\dagger}$$

where $U_\theta=UD_\theta$, $U$ is a fixed $d\times d$ unitary matrix,  and $D_\theta=\mathrm{diag}(e^{i\theta_1},...,e^{i\theta_d})$, a diagonal unitary matrix with random phases $\{\theta_j\}_{j=1}^{d}$ being i.i.d. random variables, distributed according to a probability measure $\mu$ on $(-\infty, \infty)$.
\vskip 0.2in



To define the random phases properly, we introduce the probability space $(\Theta, \mathcal{F}, \mathbb{P})$ where $\Theta=\{\theta |\theta=(\theta_1, \theta_2,..., \theta_d) \in \mathbb{R}^d\}$ ($\mathbb{R}=(-\infty,\infty)$), $\mathcal{F}$ is the $\sigma$-algebra generated by cylinders of Borel sets, and $\mathbb{P}=\otimes_{j=1}^d \mu$, where $\mu$ is a probability measure on $\mathbb{R}$ such that $\mathrm{Var}(e^{i\theta_j})\ne 0$ for $j=1, 2, ..., d$. We denote expectation values with respect to $\mathbb{P}$ by $\mathbb{E}$.

Choosing $U_\theta$ randomly according to the distribution $\mathbb{P}$ defined above, we obtain a unitary conjugation channel-valued variable 

$$\Theta\rightarrow \mathcal{L}(\mathcal{M}_d(\mathbb{C}))$$
$$\theta \mapsto \Phi_{U_\theta}$$

For a fixed $\theta$, one can check that the spectrum of $\Phi_{U_\theta}$ is
$$\mathrm{spec}(\Phi_{U_\theta}) = \{\lambda_1\overline{\lambda_2}|\lambda_1, \lambda_2\in \mathrm{spec}(U_{\theta})\}.$$

For instance, let us choose $U=\mathbb{I}$, one gets the spectrum of $\Phi_{U_\theta}$ is $\{e^{i(\theta_j-\theta_k)}|j,k=1, 2, ..., d\}$.

For this type of quantum channel-valued variable, to reason about its behavior, we are interested in its expectation with respect to the probability measure $\mathbb{P}$, which is defined by 
$$\mathbb{E}(\Phi_{U_\theta})(\rho)=\mathbb{E}(U_\theta \rho U^{\star}_{\theta})$$

For the sake of convenience, we call this quantum channel as ``mean unitary conjugation channel", abbreviated to ``MUCC". It can be checked that $\mathbb{E}(\Phi_{U_\theta})$ is a bistochastic quantum channel (unital and trace preserving). One may want to ask: what are the properties of a {\it generic} quantum channel? Our focus is on its spectral properties. The main results on its spectral properties are presented in theorem 1 and its corollary.

\vskip 0.2in
Let $\mathcal{C}$ be the set of all quantum channels that have $1$ as a simple eigenvalue and all other eigenvalues are contained in the open unit disc. A relevant result is recorded here for future reference.  

\vskip 0.2 in

Proposition 1. (see, for instance, \cite{NP2012})\,\, Consider a quantum channel $\Phi \in \mathcal{C}$. Then, for all density matrices $\rho_0\in \mathcal{M}_d^{1,+}(\mathbb{C})$
$$\lim_{n\rightarrow \infty}\Phi^n(\rho_0)=\rho_{\infty}$$
where $\rho_{\infty}$ is the unique invariant state of $\Phi$.

\vskip 0.2 in

Before showing main results of this article, we present two lemmas pertain to the behavior of unitary random variables.
\vskip 0.2in
Lemma 1.\,\, If $\int \lambda_\theta d\mathbb{P}=1$ with $|\lambda_{\theta}|=1$ a.s., then $\lambda_\theta=1$ a.s..

\vskip 0.2in

This is a well-known fact. We omit the poof of it.


\vskip 0.2in

Lemma 2.\,\, Suppose that $\{X_j\}_{j=1}^{d}$ are i.i.d. random variables, and each variance $\mathrm{Var}(X_j)\ne 0$. If $\sum_{j=1}^{d}a_jX_j=0$ a.s. with respect to the probability measure $\mathbb{P}=\otimes_{j=1}^d \mu$ , where $\{a_j\}_{j=1}^{d}$ are complex-valued scalars, then $a_j=0$ for each $j$ from $1$ to $d$.

\vskip 0.2in
Proof.\,\, Note that for the i.i.d. family $\{X_j\}$, $0=\mathrm{Var}[\sum_ja_jX_j]=\sum_ja_j^2\mathrm{Var}[X_j]$. Since the common variance is non-zero, all the $a_j$ must be null.\qed


\vskip 0.2in

Theorem 1. \,\,Let $U$ be a fixed unitary matrix of size $d$, and let the phases $\{\theta_j\}_{j=1}^{d}$ be i.i.d. random variables, distributed according to a probability measure $\mu$ on $(-\infty, \infty)$ as described above. Then the following two assertions hold.

\begin{enumerate}
  \item When each element of $U$ is non-zero, then $\mathbb{E}(\Phi_{U_\theta})\in \mathcal{C}$. It is noted that $d^{-1}\mathbb{I}$ is the unique invariant state of $\mathbb{E}(\Phi_{U_\theta})$.
  \item When each element on the main diagonal of $U$ is non-zero, then $1$ is the only eigenvalue of $\mathbb{E}(\Phi_{U_\theta})$ on the unit circle and all other eigenvalues are contained in the open unit disc. 
  
In particular, when $U$ is a diagonal matrix, then the eigenspace of $\mathbb{E}(\Phi_{U_\theta})$ is spanned by $\{\mathrm{diag}(1,0,...,0),...,\mathrm{diag}(0,...,0,1)\}$.
\end{enumerate}

Proof.\,\, It is evident that $1$ is one of its eigenvalues and the completely mixed density operator $d^{-1}\mathbb{I}$ is an eigenvector corresponding to $1$. Now let's suppose that $\lambda$ is one of its eigenvalues with unit absolute value, and $\rho$ is a corresponding eigenvector, i.e., $\mathbb{E}(\Phi_{U_\theta})(\rho)=\lambda \rho$ with $|\lambda|=1$. We have the following identities and inequalities:

\begin{eqnarray}
|\langle \rho, \lambda \rho\rangle|=\|\rho\|^2=|\langle \rho, \int U_{\theta}\rho U^{\star}_{\theta}d \mathbb{P}\rangle|=|\int \langle \rho, U_\theta \rho U^{\star}_\theta \rangle d\mathbb{P}| \nonumber \\
\le \int |\langle \rho, U_\theta \rho U^{\star}_\theta \rangle |d\mathbb{P}\le \|\rho\|^2 \label{essential_1}
\end{eqnarray}
Here the inner product is the {\em Hilbert-Schmidt} inner product, defined by $\langle A, B\rangle=\mathrm{tr}(A^{\dagger}B)$, the norm is induced by this inner product, i.e., $\|A\|=\sqrt{\mathrm{tr}(A^{\dagger}A)}$.

Now one can observe that both inequalities in Eq.(\ref{essential_1}) are actually equalities. In particular, the second equality is true if and only if $U_\theta \rho U^{\star}_\theta$ and $\rho$ are linearly dependent a.s. w. r. t. $\mathbb{P}$, by the Cauchy-Schwarz inequality. Therefore we may assume that $U_\theta \rho U^{\star}_\theta=\lambda_\theta \rho$ with $|\lambda_\theta|=1$ a.s.. Note that 

\begin{eqnarray}
\lambda\|\rho\|^2=\langle \rho, \lambda \rho\rangle=\langle \rho, \int U_\theta \rho U^{\star}_{\theta}d\mathbb{P}\rangle=\int\langle \rho, U_\theta \rho U^{\star}_{\theta}\rangle d\mathbb{P}=\|\rho\|^2\int \lambda_\theta d\mathbb{P}
\end{eqnarray}
which implies that $\lambda=\int \lambda_\theta d\mathbb{P}$, thus one can deduce that $\lambda_\theta=\lambda$ a. s. by Lemma 1.

In summary, as shown from the preceding reasoning, $\mathbb{E}(\Phi_{U_\theta})(\rho)=\lambda \rho$ with $|\lambda|=1$  implies $U_\theta \rho U^{\star}_\theta=\lambda \rho$ a.s., which is equivalent to $U_\theta \rho=\lambda \rho U_\theta$ a.s..

Before we proceed further, we announce that all arguments below are based on Lemma 2.


In what follows we aim at proving two facts about the aforesaid $\lambda$ and $\rho$: (1) $\lambda=1$; (2) $\rho$ is a diagonal matrix. It should be stressed that the only condition imposed here on the channel $\mathbb{E}(\Phi_{U_\theta})$ is the non-nullity of entries on the main diagonal of $U$, namely, $u_{kk}\ne 0$ for $k=1, 2, ..., d$ ( the two cases described in the theorem satisfies this condition ). Let us choose $(L_{kj})$ and $(R_{kj})$ to denote matrices $U_\theta \rho$ and $\lambda \rho U_\theta$, respectively, then $L_{kj}=\sum_lu_{kl}\rho_{lj}e^{i\theta_l}$, $R_{kj}=\lambda\sum_lu_{lj}\rho_{kl}e^{i\theta_j}$. For each $k\in \{1, 2,..., d\}$, using $u_{kk}\ne 0$, one can deduce that $\rho_{kj}=0$ for $j\ne k$. This has shown that the eigenvector $\rho$ actually is a diagonal matrix, which can be written as $\rho=\mathrm{diag}(\rho_{11},...,\rho_{dd})$. Note that $\rho\ne 0$, one may assume that $\rho_{11}\ne 0$ without loss of generality. Then $\lambda=1$ immediately follows from $L_{11}=R_{11}$.


Case 1.\,\, $U\ne 0$, meaning each element of the matrix $U$ is non-zero. Under this circumstance, the equalities $L_{k1}=R_{k1}$ for $k=1, 2, ..., d$ imply that $\rho_{11}=\rho_{22}=...=\rho_{dd}$. Then one may choose $\rho=d^{-1}\mathbb{I}$.


Case 2.\,\,Considering the preceding arguments, we only need to prove that any diagonal matrix $\rho$ can be an eigenvector of $\mathbb{E}(\Phi_{U_\theta})$ corresponding the unit eigenvalue when $U=\mathrm{diag}(u_{11},u_{22},...,u_{dd})$. This is obviously true due to the fact that multiplication is commutative for diagonal matrices.\qed


\vskip 0.2 in

To illustrate that the non-nullity of entries on the main diagonal of $U$ is just a sufficient condition, for that $1$ is the only eigenvalue of $\mathbb{E}(\Phi_{U_\theta})$ on the unit circle and all other eigenvalues are contained in the open unit disc. Let us take a look at a specific channel $\mathbb{E}(\Phi_{U_\theta})$ with $U$ defined by

\begin{equation}
 \sigma_1= \left[\begin{array}{cc}
0& 1\\
1& 0
\end{array}\right]. \label{}
\end{equation}
 
Without much difficulty, one can show that both $1$ and $-1$ are eigenvalues of $\mathbb{E}(\Phi_{U_\theta})$, and their respective eigenspaces are spanned by $2^{-1}\mathbb{I}$ and $\sigma_3$. Here, $\sigma_3$ is defined by 

\begin{equation}
 \sigma_3= \left[\begin{array}{cc}
1& 0\\
0& -1
\end{array}\right]. \label{}
\end{equation}

\vskip 0.2 in

Applying the aforesaid Proposition 1 and Theorem 1 together leads to the following result on the asymptotic state of the system after $n$ iterations.

\vskip 0.2in
Corollary \,\, If each element of $U$ is nonzero, then $\mathbb{E}(\Phi_{U_\theta})$ has
an unique invariant state $d^{-1}\mathbb{I}$ and for all density matrices $\rho_0$,

$$\lim_{n\rightarrow \infty}(\mathbb{E}(\Phi_{U_\theta}))^n \rho_0=\frac{\mathbb{I}}{d}$$
 
\vskip 0.2in

Even if a limiting state $[\Phi_{U_\theta}]^n\rho_0$ does not necessarily exist in the usual sense, we still might want to probe the possibility of a "limiting state" $\rho_{\infty}$ in the sense of Ces`aro:

\begin{equation}
\rho_{\infty}=\lim_{n\rightarrow \infty}\frac{[\Phi_{U_\theta}]\rho_0+[\Phi_{U_\theta}]^2\rho_0+...+[\Phi_{U_\theta}]^n\rho_0}{t}
\end{equation}
In terms of this generalized sense of "limiting state," it turns out that every quantum channel converges.

\vskip 0.2in

Proposition 2. \,\, Let $U$ be a unitary matrix without a zero element. Then, almost surely,
\begin{equation}
\lim_{n\rightarrow \infty}\frac{[\Phi_{U_\theta}]\rho_0+[\Phi_{U_\theta}]^2\rho_0+...+[\Phi_{U_\theta}]^n\rho_0}{t}=\frac{\mathbb{I}}{d}
\end{equation}

Proof. Assume that $\Phi_{U_\theta}\rho=\lambda \rho\Phi_{U_\theta}$, a.s., one can derive that $\lambda=1$ and  $\rho=\frac{\mathbb{I}}{d}$ via the same reasoning as employed in the proof of Theorem 1. Then the conclusion follows upon applying Theorem 7 in \cite{LP2012}.\qed


\vskip 0.2in
It is noteworthy that a similar result for the random quantum channels with randomness defined by Haar-distributed unitaries can be found in \cite{NP2012}.

\vskip 0.2 in
By Theorem 1 and its Corollary, a large number of iterations of a quantum channel can map a state into one which is very ``close" to the maximal mixing state. In what follows, we present a model of unitary conjugation channels with an arbitrary size, a cycle of two iterations maps exactly an any state $\rho$ into the maximal mixing density $\frac{1}{2^k}\mathbb{I}$.

Let $H$ be Hadamard matrix, which is given by 

\begin{equation}
 H= \frac{\sqrt{2}}{2}\left[\begin{array}{cc}
1& 1\\
1& -1
\end{array}\right]. \label{H}
\end{equation}

Consider the unitary conjugation channel $\mathbb{E}(\Phi_{U_\theta(k)})$ where 

$U_{\theta}(k)=H^{\otimes k}\mathrm{diag}(e^{i\theta_1},...,e^{i\theta_{2^k}})$, and the random phases $\{\theta_j\}_{j=1}^{2^k}$ are assumed to be i.i.d. random variables, distributed uniformly on $(-\pi, \pi)$. In these settings for $\mathbb{E}(\Phi_{U_\theta(k)})$, one can claim that two iterations of the channel maps any density $\rho_0$ into the maximal mixing density $\frac{1}{2^k}\mathbb{I}$, i.e., 

$$(\mathbb{E}(\Phi_{U_\theta(k)}))^2 \rho_0=\frac{\mathbb{I}}{2^k}.$$

 
Here is a brief outline of justification for this claim: Since $\mathbb{E}(e^{i\theta})=0$, $\mathbb{E}(\mathrm{diag}(e^{i\theta_1},...,e^{i\theta_{2^k}})\rho_0 \mathrm{diag}(e^{-i\theta_1},...,e^{-i\theta_{2^k}}))=\mathrm{diag}(\rho_{11},...,\rho_{2^k2^k})$ where $\rho_0=(\rho_{kj})_{2^k\times2^k}$. Then $\mathbb{E}(\Phi_{U_\theta(k)})\rho_0=H^{\otimes k}\mathrm{diag}(\rho_{11},...,\rho_{2^k2^k})H^{\otimes k}$, which is the output of the first iteration and is denoted by $\rho_1$, this is a matrix with all entries on the main diagonal being the constant $1/2^k$. After performing second iteration of the channel $\mathbb{E}(\Phi_{U_\theta(k)})$ on $\rho_1$, one can see that the output is the maximal mixing density $\frac{1}{2^k}\mathbb{I}$. 

\vskip 0.2in

Having examined the spectral properties of an MUCC, now one may want to know if an MUCC could be described by discrete phases? In other words, are discrete phases sufficient to achieve quantum channel $\mathbb{E}(\Phi_{U_\theta})$? If yes, then what is an {\em operator-sum representation} \cite{NC2000} for an MUCC? In what follows, we shall present an operator-sum representation for an MUCC with specified distributed phases. 

\vskip 0.2in
Theorem 2. \,\,When random phases $\{\theta_j\}_{j=1}^{d}$ are i.i.d. random variables, distributed uniformly on $(-\pi, \pi)$, then $\mathbb{E}(\Phi_{U_\theta})$ can be defined by Kraus operators $\frac{1}{\sqrt{2^d}}U\mathrm{diag}(e^{i\theta_1},...,e^{i\theta_d})$ where $\theta_n$ is randomly chosen from $\{-\frac{\pi}{2},\frac{\pi}{2}\}$, i.e., an {\em operator-sum representation} for $\mathbb{E}(\Phi_{U_\theta})$ is given by 
$$\mathbb{E}(\Phi_{U_\theta})(\rho)=\frac{1}{2^d}\sum_{\theta_1,\theta_2,...,\theta_d\in \{-\frac{\pi}{2},\frac{\pi}{2}\}}U\mathrm{diag}(e^{i\theta_1},...,e^{i\theta_d})\rho \mathrm{diag}(e^{-i\theta_1},...,e^{-i\theta_d})U^{\dagger}.$$

Proof.\,\,Under the assumption of the theorem, $\mathbb{E}(e^{i\theta})=0$, which implies that $U\mathbb{E}(D_\theta \rho D^{\dagger}_\theta)U^{\dagger}=\mathbb{E}(\Phi_{U_\theta})(\rho)=\mathbb{E}(U_\theta \rho U^{\star}_{\theta})$. It suffices to only justify that 

\begin{equation}
\mathbb{E}(\Phi_{U_\theta})(\rho)=\frac{1}{2^d}\sum_{\theta_1=-\frac{\pi}{2},\frac{\pi}{2}},...,\sum_{\theta_d=-\frac{\pi}{2},\frac{\pi}{2}}\{\sum_n e^{i\theta_n}|\epsilon_n\rangle\langle \epsilon_n|\rho\sum_n e^{-i\theta_n}|\epsilon_n\rangle\langle \epsilon_n|\} \label{mean}
\end{equation}

is true when $U=\mathbb{I}$. Let us assume that $\rho=(x_{jl})$ in an orthonormal basis $\{|\epsilon_n\rangle\}$, note that $\mathbb{E}(e^{i\theta})=0$, it is easy to verify that $\mathbb{E}(\Phi_{U_\theta})(\rho)=\sum_{n}x_{nn}|\epsilon_n\rangle \langle \epsilon_n|$. Since  
$$\sum_n e^{i\theta_n}|\epsilon_n\rangle\langle \epsilon_n|\rho\sum_n e^{-i\theta_n}\epsilon_n\rangle\langle \epsilon_n|=\sum_j\sum_le^{i(\theta_j-\theta_l)}x_{jl}|\epsilon_j\rangle \langle \epsilon_l|$$ with $e^{i(\theta_j-\theta_l)}=1$ or $-1$ when $\theta_j$ and $\theta_l$ are $-\frac{\pi}{2}$ or $\frac{\pi}{2}$, the right-hand side of Eq. (\ref{mean}) is $\sum_{n}x_{nn}|\epsilon_n\rangle \langle \epsilon_n|$. Therefore, the conclusion of the theorem follows.\qed

\vskip 0.2in

Theorem 2 shows that the mean unitary conjugation channel (MUCC) with some specific distributed phases, can be characterized by {\em random unitary channels} \cite{NC2000}. It would be interesting to attempt to discretize an MUCC with general distributed phases.

\vskip 0.2in


\vskip 0.2in



\begin{thebibliography}{0}

\bibitem{KBH2006}

Kosk J., Buzek V., and  Hillery M.: {\it Quantum walks with random phase shifts}, Phys. Rev. A 74, 022310(2006).


\bibitem{JM2010}

Joye A., Merkli M.: {\it Dynamical localization of quantum walks in random environments}, J Stat Phys (2010)140: 1025-1053.

\bibitem{HW2008}
Hayden P., Winter A.: {\it  Counterexamples to the maximal p-norm multiplicativity conjecture
for all $p>1$}, Comm. Math. Phys. 284(1):263-280, 2008.

\bibitem{H2009}

Hastings M. B.: {\it Superadditivity of communication capacity using entangled inputs}, Nature Physics 5, 255 - 257 (2009).

\bibitem{CN2010}

Collins B., Nechita I.: {\em Random quantum channels I: graphical calculus and the Bell state phenomenon}, Communications in Mathematical Physics 297, 2 (2010) 345-370.


\bibitem{NP2012}

Nechita I., Pellegrini C.: {\it Random repeated quantum interactions and random invariant states}, Probability Theory and Related Fields, Volume 152, Issue 1-2, pp 299-320(2012).






\bibitem{LP2012}

Liu C., Petulante N.: {\em On limiting distributions of quantum Markov chains}. arXiv:1010.0741 (2010). International Journal of Mathematics and Mathematical Sciences, Volume 2011 (2011), Article ID 740816.


\bibitem{NC2000}

Nielsen M., Chuang I.: {\it Quantum Computation and Quantum Information} (Cambridge University Press, Cambridge, 2000).

\end{thebibliography}
\end{document}